\documentstyle[aps,prl,epsfig,floats]{revtex}
\begin{document}
\draft
\twocolumn[\hsize\textwidth\columnwidth\hsize\csname
@twocolumnfalse\endcsname
\title{Effective Lorentz Force due to Small-angle Impurity
Scattering: Magnetotransport in
High-$T_c$ Superconductors}
\author{C.M. Varma}
\address{Bell Laboratories, Lucent Technologies
Murray Hill, NJ 07974}
\author{Elihu Abrahams}
\address{Center for Materials Theory, Serin Physics Laboratory, Rutgers
University, Piscataway, NJ 08854-8019}
\date{\today}
\maketitle
\begin{abstract}

We show that a scattering rate which varies with angle around the Fermi surface
has the same effect as a periodic Lorentz force on
magnetotransport  coefficients. This effect, together with
the marginal Fermi liquid inelastic scattering rate gives
a quantitative explanation of the temperature dependence and the
magnitude of the observed Hall effect and magnetoresistance
with just the measured
zero-field resistivity as input.
\end{abstract}

]

The temperature dependence of the transport properties in
the normal state of high-temperature superconductors \cite{review} are
unlike Landau Fermi liquids.  Most of the observed
anomalies near the composition for the highest $T_c$ follow
from the hypothesis of a scale-invariant
fluctuation spectrum \cite{mfl}, characteristic of a quantum critical
point, which is a function of $\omega/T$ and which has
negligible momentum dependence over most of the
Brillouin zone. One of the principal predictions
of this hypothesis is that at low energies the inelastic part of
the single-particle relaxation rate has the marginal Fermi liquid
(MFL) form $\lambda T$ with coefficient $\lambda$ having negligible
momentum dependence either along or perpendicular to the Fermi surface.
This prediction has been verified in detail in recent angle-resolved
photoemission (ARPES) measurements \cite{arpes}.

However, the observed anomalies in the  magnetotransport
\cite{ong,hwang,drew} do not follow from the MFL
hypothesis.  Various explanations \cite{halltheories,KSV} have been
advanced for these anomalies, none
of which are independently supported by other experiments,
and in some cases are in conflict with photoemission experiments.

The ARPES experiments \cite{arpes} have revealed, besides the inelastic
contribution to the single-particle self energy, an elastic
contribution (independent of temperature or frequency)
which is angle-dependent around the Fermi surface, increasing by
 about a factor of 4 from the $(\pi,\pi)$ to the $(\pi,0)$
directions.
Thus the total self energy at low frequency has an imaginary part
of the form
\begin{equation}
{\rm Im}\: \Sigma ({\vec k},T)\  =\  \Gamma_{0}(\hat k)
+\lambda T,
\end{equation}
The anisotropic elastic part, $\Gamma_0(\hat k)$, has been ascribed to
small-angle scattering from dopant impurities lying between the
Cu-O planes \cite{pnas}.  If
$d$ is the characteristic distance of such impurities from
a Cu-O plane the electron scattering at a point $\hat k$
on the Fermi-surface is confined to
small momentum transfers $\delta k \leq d^{-1}$.
Then, on the Fermi surface, the scattering
rate at a point $\hat k$ is proportional to
$\delta k$, to the local density of states and to the forward scattering
matrix element at $\hat k$. The latter two quantities can be anisotropic especially
near a crossover or a transition to an anisotropic pseudogap
state. We do not address the sources of anisotropy here and simply take
$\Gamma_0(\hat k)$ from experiment and follow the consequences.

In a planar lattice of square symmetry, the anisotropic scattering rate
$\Gamma_0 (\hat {k})$, has a four-fold symmetry as one goes around the
Fermi surface. Since it is largest in the $(\pi,0)$ directions
\cite{pnas}, a non-equilibrium particle distribution is skewed away
from the region of the Fermi surface near the
$(\pi,0)$ directions toward the $(\pi, \pi)$ directions.
We show that this acts as an effective magnetic field, perpendicular to the
planes which changes sign at every multiple of $\pi/4$ in going around the
Fermi surface. Together with the MFL scattering rate, such a skewed
distribution gives an anomalous contribution to the
magnetotransport properties with the experimentally observed
temperature dependences. Using just the measured dc resistance,
we can explain the observations quantitatively.
We also explain the observed variation in the anomalous
magnetotransport properties due to impurities like Zn
which replace Cu in the plane and cause angle-independent
($s$-wave) scattering.

The skewed non-equilibrium distribution is apparent from
the (formal) solution of the linearized Boltzmann equation.
Let $g ({\bf k})$ be the deviation from the equilibrium distribution
$f_0 ({\bf k})$ due to the applied uniform steady state electric
and magnetic fields ${\bf E}$ and ${\bf B}$.  It is formally given
by \cite{KSV}
\begin{equation}
g({\bf k})=e\sum _{\bf k'}\  A^{-1}_{{\bf k},{\bf k}'}\left[{\bf E}
\cdot {\bf v}_{{\bf k}'}\left( -\frac{\partial f_0}{\partial
\epsilon _{{\bf k}'}}
\right)\right],
\end{equation}
where
\begin{equation}
A_{{\bf k},{\bf k}'}=\left[{1\over \tau({\bf k}) }
\  +{e\over \hbar c}\:
{\bf v}_{\bf k}\times {\bf B}\cdot \nabla _{\bf k}\right]\delta_
{{\bf k},{\bf k}'}-C_{{\bf k},{\bf k}'}.
\end{equation}
Here $C_{{\bf k},{\bf k}'}$ is the ``scattering-in''
term in the collision operator for the Boltzmann  equation
and $ 1/\tau({\bf k})
=\sum _{{\bf k}'}C_{{\bf k},{\bf k}'}$, is
the ``scattering-out'' term equal to the
single-particle
relaxation rate.
From Eqs.\ (3,4) It is evident that the distribution
$g({\bf k})$ is not merely determined by
the energy of the state ${\bf k}$
but also by the anisotropy of the scattering. The
distribution is depleted in directions of large net scattering
and augmented in directions of small net scattering.

We shall calculate the conductivities for
$\omega \ll T$ using the single-particle scattering rate of Eq.\ (1).
The
conductivity tensor is
\begin{equation}
\sigma ^{\mu \nu }=e^{2}\sum _{{\bf k},{\bf k}'} \:
 v _{\mu ,{\bf k}}\  A_{\bf k,\bf k' }^{-1} \:
v_{\nu ,{\bf k}}\left(-{\partial f_{0}\over \partial
\epsilon _{\bf k '}} \right).
\end{equation}
We expand $A^{-1}$ from Eq.\ (3) in powers of $\bf B$ \cite{KSV}:
\begin{eqnarray}
A^{-1}={\cal T} &-&{\cal T} ({\bf v}\times {\bf B\cdot \nabla })\,
{\cal T} \nonumber \\
&+&{\cal T}
({\bf v\times B\cdot \nabla})\,{\cal T}\,({\bf v\times B\cdot
\nabla })\,{\cal T},
\end{eqnarray}
where, from Eq.\ (2.12) of Ref.\ \cite{KSV}, it follows that
\begin{equation}
{\cal T} ({\bf k}, {\bf k}^\prime ) = \tau(\bf k)\delta_{{\bf k},
{\bf k}'}
 + \sum_{{\bf k}''} \tau(\bf k)C_{{\bf k},{\bf k}''}
{\cal T} ({\bf k}'', {\bf k}^\prime ).
\end{equation}
The first term in $A^{-1}$ gives the dc conductivity,
$\sigma^{xx}$ etc., the second the Hall conductivity $\sigma^{xy}$
and the
magnetoconductivity may be calculated from the third term.

Let $\theta,\, \theta'$ be the angles of ${\bf k,\, k}'$ with
respect to the $k_x$ axis in reciprocal space. The use of the
Boltzmann formulation and the presence of $-\partial f_0/\partial
\epsilon$ in Eq.\ (2) essentially restricts all variables to the
Fermi surface; it implies that all the scattering is quasielastic.
We take $C_{{\bf k},{\bf k}'}$ to be composed of three
parts:
\begin{equation}
C_{{\bf k},{\bf k}'}=  2 \pi \delta (\epsilon_{\bf k} -
\epsilon_{{\bf k}'})
\left(
|U_M|^2 + |u_{i}(\theta ,\theta^{\prime})|^2 + |U_{i}|^2\right).
\end{equation}
The first contribution, due to MFL, is parameterized
by an angle independent quasielastic scattering proportional to
$T$. The
second and the third terms are due to small angle
and large angle elastic impurity scattering respectively.
The effect of  $U_{i}$ is simply to add a temperature independent scattering
rate to the MFL scattering rate. $u_i (\theta, \theta^\prime )$ is a symmetric
function of
$\theta$ and $\theta^{\prime}$ sharply peaked at
$\theta'=\theta$ with a small characteristic width
$\theta_c(\theta)$.
Then to order $\theta_c^3$ the single particle
scattering rate is
\begin{eqnarray}
\tau^{-1} (\theta ) & = & \tau_M^{-1}+ \tau _{i}^{-1}(\theta ) \\
\tau _{i}^{-1}(\theta ) & = & \theta_c|u_i(\theta, \theta )|^2
N(\theta ) \nonumber \\
& &+ \left(\frac{\theta _{c}^{3}}{24}\right){\partial ^{2}\over
\partial \theta'^{2}}
\left[|u_i(\theta, \theta' )|^2 N(\theta' )\right]_{\theta
'=\theta},
\end{eqnarray}
where $N(\theta)$ is the density of states per radian at the Fermi surface
at angle $\theta$ and $\tau_M^{-1}$ is the sum of the MFL scattering rate
proportional to $T$ and the large angle impurity scattering rate.

Given the quasielastic $C_{{\bf k},{\bf k}'}$, one may
integrate over the energy variable $\epsilon_{\bf k'}$ (i.e.
perpendicular
to the Fermi surface) in Eq.\ (4). To evaluate the conductivities,
it is convenient to define the operator
$\hat{\tau}$ by
\begin{eqnarray}
\hat{\tau}(\theta )v_{\alpha }(\theta )(-\partial f_0/\partial
\epsilon_{{\bf k}})
&\equiv & \sum_{{\bf k}'}{\cal T}_{{\bf k,k}'} v_{\alpha,{\bf k}'}
(-\partial f_0/\partial\epsilon_{{\bf k}'}) \nonumber\\
& = & {1\over 2\pi}
\int d\theta' N(\theta')
{\cal T}(\theta ,\theta ')v_{\alpha }(\theta '),
\end{eqnarray}
where the $v_{\alpha}$ are  the
Fermi surface velocity components.

From Eqs. (6,7) it follows that
\begin{equation}
\hat{\tau}(\theta )v_{\mu }(\theta )\ =\
 \tau (\theta ) v_{\mu}
(\theta )+ {\hat {\rm t}}v_{\mu }(\theta )
\end{equation}
where
\begin{equation}
{\hat {\rm t}}v_{\mu }(\theta ) \equiv
{\tau(\theta)\over 2\pi}
\int d\theta ' N(\theta')
\  C(\theta ,\theta')\hat{\tau}(\theta')v_{\mu}(\theta'),
\end{equation}
where $C(\theta ,\theta')$ is Eq.\ (7) without the energy
$\delta$-function.

Now ${\hat {\rm t}}v_{\mu }(\theta )= 0$ for momentum-independent
scattering. This is simply the traditional result of no vertex
corrections for such scattering.
Therefore, in Eq.\ (12) we use only the small-angle scattering
part in $C$ and find
\begin{eqnarray}
\frac{{\hat {\rm t}}v_{\mu }(\theta )}{\tau(\theta)} &=& \left[{1 \over
\tau _{i}(\theta )}
\hat{\tau}(\theta) v_{\mu }(\theta)\right] \nonumber \\
&-& \left({\theta_c^3\over 24}\right){\partial^{2}
\over
\partial\theta'^{2}}
\left [N(\theta')~|u_i(\theta, \theta' )|^2 \right]_{\theta
'=\theta}\hat{\tau}(\theta) v_{\mu }(\theta ) \nonumber\\
&+&\left({\theta_c^3\over 24}\right){\partial^{2}\over
\partial\theta'^{2}}\left[
N(\theta')~|u_i(\theta,\theta')|^2~
\hat{\tau}(\theta')v_{\mu }(\theta')\right]_{\theta'=\theta}
.
\end{eqnarray}
Then
\begin{equation}
{\hat\tau}v_{\mu} = \tau\left\{v_{\mu}+\left[{1\over \tau_i}
-
\left({1\over\tau_u}\right)''\right]{\hat\tau}v_{\mu}
+
\left[{1\over\tau_u}{\hat\tau}v_{\mu}\right]''\right\},
\end{equation}
where primes denote derivatives with respect to $\theta$ and
$1/\tau _u$ is the
impurity scattering contribution to the transport scattering rate and is
given, to leading non-trivial order in
$\theta_c$, by
\begin{equation}{1\over \tau _u
(\theta )} \simeq  (\theta _{c}^{2}/24){1\over \tau _{i}(\theta )}
\end{equation}
That small angle impurity scattering has much less effect
on the transport rate \cite{pnas} than on the single-particle
scattering  rate is evident in Eq. (15).

In Eq. (14), derivatives of $\hat{\tau}(\theta )$ lead to
corrections  of higher order in $\theta_c$ compared to the other terms
and may be neglected. Therefore, we have
\begin{equation}
\left({1\over \tau} - {1\over \tau_i}\,\right){\hat \tau}\,v_{\mu}= v_{\mu}
+2\left({1\over \tau_{u}}\right)'{\hat \tau}\,v_{\mu}' +
 {1\over \tau_{u}}{\hat \tau}\,v_{\mu}'',
\end{equation}

The derivatives of
the Fermi velocity components may be written as
$v_{\mu}' = \sum_{\nu}b_{\mu\nu}(\theta)v_{\nu}$.
The relevant two-dimensional Fermi surface has four-fold
symmetry. Let
$\phi(\theta)$ be the angle of the Fermi velocity $\vec v (\theta)$ with
respect to the
$k_x$ axis in the Brillouin zone. The precise form of
$b_{\mu\nu}$ depends on Fermi surface details. For example,
for the case that $|v|$ is constant, $v_x' =
-(d\phi/d\theta)v_y$ and $v_y' = (d\phi/d\theta)v_x$. It may be
verified that
$(v_x,v_y)$ form an orthogonal basis for the irreducible two-dimensional
representation of the group of the square. All other quantities which
enter into the coefficients $b_{\mu\nu}(\theta)$ and elsewhere, e.g. $|\vec
v|,\,\,d\phi/d\theta,\, d(1/\tau_{u})/d\theta,
\ldots$ transform according to the identity. We shall use these facts
below to eliminate certain integrals on symmetry grounds.

The solution to the coupled equations Eq.\ (16) has the
following form:
\begin{equation}
{\hat \tau}{\vec v}= \tau_t \,\left[{\vec v} -
(\tau_t/\tau_p)\,{\vec v}\times{\hat{\bf z}}
\right]
\end{equation}
where $\hat{\bf z}$ is the unit vector normal to the Cu-O plane.
The scattering rates $1/\tau_t$ and $1/\tau_p$ depend on the precise
shape of the Fermi surface. To simplify what follows, we take the case
that the Fermi speed is constant over the Fermi surface. In that case,
there is a single coefficient $b_{\mu\nu}$ which is just $\phi' =
d\phi/d\theta$. In addition, we keep only the leading orders in
$\theta_c$. Then, the
$\tau$s are determined by
\begin{eqnarray}
{1\over\tau_t} &=&{1\over\tau} - {1\over\tau_i} +(\phi')^2
{1\over\tau_{u}} ={1\over\tau_M} +(\phi')^2 {1\over\tau_{u}}\\
{1\over\tau_p} &=& 2~(\phi')~{d\over d\theta}\left({1\over
\tau_{u}}\right) + (\phi'')~{1\over
\tau_{u}}.
\end{eqnarray}

The structure of Eq.\ (17) shows that the deviation of the
distribution from equilibrium which is determined by ${\hat
\tau}v$ has a skew character. Thus, there is an effective
Lorentz force which rotates the carrier distribution into the
directions of weak small-angle scattering. Because
$d(1/\tau_{u})/d\theta$ changes sign at each $\theta =
n\pi/4$,  carriers moving on the Fermi surface experience an
effective magnetic field which changes sign eight times around
the Fermi surface. This has important consequences for
$\sigma^{xy}$.

Knowing $\hat\tau\,v$, we evaluate
$\sigma^{xx}$ from Eqs.\ (4,5,11):
\begin{eqnarray}
\sigma^{xx} &=& e^2\sum_{{\bf k}{\bf k}'} v_x({\bf k}) {\cal
T}_{{\bf k}{\bf k}'} v_x({\bf k}') (-\partial
f_0/\partial\epsilon') \nonumber \\ &=& (e^2/2\pi)\int d\theta
N(\theta) v_x(\theta) [{\hat{\tau}}(\theta)v_x(\theta)].
\end{eqnarray}
Using Eq. (17), we obtain
\begin{equation}
\sigma^{xx} =(e^2/2\pi)\int d\theta N(\theta) [v_x(\theta)]^2
\tau_t(\theta).
\end{equation}

For $\sigma_{xy}$, we use Eq.\ (17) again and find
\begin{equation}
\sigma^{xy}={e^3B\over 4\pi^2c}\int d\theta\, \tau_t(v_x -
{\tau_t\over\tau_p}v_y){\partial\over\partial\theta}\left[\tau_t(v_y +{\tau_t\over
\tau_p}v_x)\right].
\end{equation}

Finally, we get, to leading orders in $\theta_c$,
\begin{equation}
\sigma^{xy}={e^3B\over 4\pi^2c}\int d\theta  [\phi'\, \tau_t^2 v_x^2
+ \tau_t^3\, (\partial \tau_{p}^{-1}/\partial\theta)\, v_x^2].
\end{equation}

Eqs. (21,23) are our main results. From them we may determine
the Hall angle $\Theta_H= {\rm
tan}^{-1}(\sigma^{xy}/\sigma^{xx})$. We find
\begin{equation}
\tan\Theta_H  =  \tan(\Theta_H)_0 + \omega_H \,\overline
{\tau^2_t(\tau_p^{-1})'}
\end{equation}
where the first term is the customary contribution
and in the new contribution which we have found, the overbar represents
an appropriate average over the Fermi surface according to Eqs.
(21,23). Our principal finding is that new contribution has the
temperature dependence of the resistivity relaxation time $\tau_t$
squared.

\begin{figure}[htb]
\centering
\epsfxsize=\linewidth
\epsffile{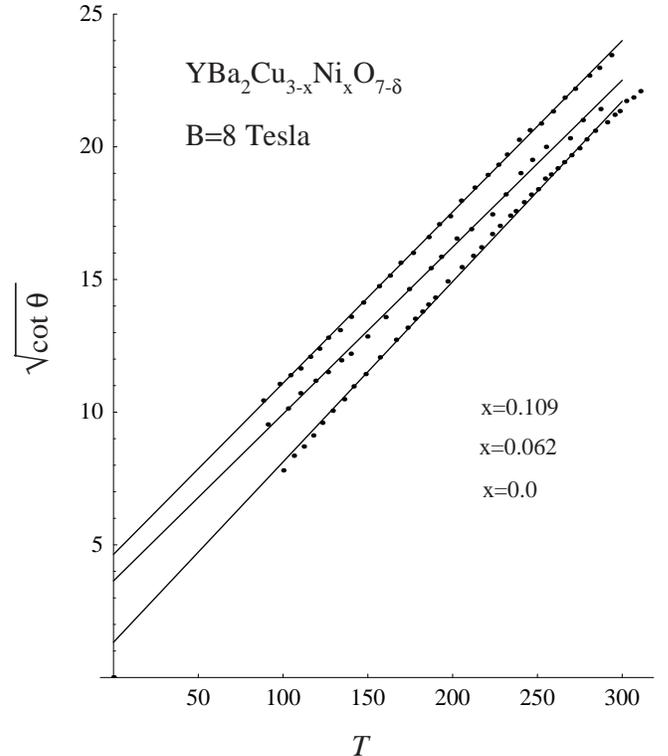}
\caption{Hall angle data for three concentrations of Zn in
YBCO. The straight lines are least-squares fits to the data (replotted
from Chien {\it et al} [4]) for
$\sqrt{\cot\theta_H(T)}$ vs $T$.}
\label{sp_heat}
\end{figure}

The customary contribution depends on the details of the band structure.  For
YBCO near optimum composition, band-structure calculations \cite{pick} find for
the Hall constant a value which is about five times smaller than what is
actually measured.  The customary contribution to $\tan(\Theta_H)$ is then at
least five times smaller than the measured value and is probably even smaller
due to Fermi-liquid effects.  Then with neglect of the customary term, our
conclusion is that $(\cot\Theta_H)^{1/2}$ should be proportional to
$\tau^{-1}_t$ which is independently measured by the zero-field resistivity.  As
seen in Eq.\ (18), $\tau^{-1}_t$ contains an impurity-independent part with MFL
linear $T$ dependence and a $T$-independent part depending on impurity content.
Fig.\ 1 shows the comparison for samples with varying concentration, $x$, of Zn
impurities in an applied magnetic field of 8 Tesla.  The straight lines are
least-square fits to the data from Chien {\it et al} [4].  The data is
consistent with a slope independent of $x$, as predicted.  The increase of the
intercept with $x$ is due to the increased large-angle scattering from in-plane
impurities.  For well-prepared samples, the ratio of resistivity (hence of
$1/\tau_t$) at 100 K to the 0 K extrapolated value is usually about 10.  This is
consistent with the present analysis which gives $\sim 7$ for the corresponding
ratio for $[\cot\Theta_H(T)]^{1/2}$ from the data. 
The magnitudes of the various terms
which enter the transport coefficients depend very sensitively on the shape of
the Fermi surface and the precise variation with angle of the impurity
scattering rate $1/\tau_i$. To make further estimates, we adopted a model Fermi
surface similar
to the measured one. We estimate the small-angle parameter $\theta_c$:  The
extrapolated resistivity at $T=0$K is determined by $1/\tau_{u} =
(\theta_c^2/24)(1/\tau_i)$ and that at $T=100$K is determined by
$1/\tau_M(100$K).  Using the ARPES-measured $1/\tau_i$ and $1/\tau_M$ and a
typical resistivity ratio $\rho(100$K)/$\rho(0$K) of 9, we find
$\theta_c^2/24 \approx .01$.  We use the measured $1/\tau_M$ and the
above-determined $1/\tau_{u}$ to compare the new contribution to $\sigma^{xy}$
to the conventional one - i.e.  the ratio of the integrals of the second to the
first term in square brackets in Eq.\ (23).  For the model Fermi surface, we
found the new contribution to be almost two orders of magnitude larger that
the conventional
piece, even without invoking the band-structure result mentioned earlier.  We
have thus given a demonstration-in-principle of our principal finding.  Since
the effects of the anisotropic small-angle scattering can be quite large, it is
natural to ask whether higher order terms in $\theta_c^2/24$ are important.
Using the same model, we find them to be less that 3\% of the leading term.

Ong \cite{ongmagneto} has derived a relation between the
magnetoresistance and moments of the fluctuations of the Hall
angle over the Fermi surface in two dimensions. This relationship
continues to hold in the present case.  For small Hall angle, an order of
magnitude estimate is
\begin{equation}
(\Delta \rho /\rho)(B) \sim \Theta^2_H,
\end{equation}
Using similar arguments to those of the previous paragraph, our result for the
Hall
angle gives the measured temperature dependence of the
magnetoresistance\cite{ong} as well as order of magnitude numerical agreement.

We extend the theory to low frequencies ($\omega\tau_t \ll 1$) by means of the
replacement $1/\tau_M \rightarrow i\omega+\tau^{-1}_M$ so that
$\cot\theta_H \propto
(i\omega + \tau_t^{-1})^2$. The observed low-frequency behavior of the
real and imaginary parts of the Hall angle
\cite{drew} then follow.

Two properties which follow from the
MFL phenomenology and experiment are crucial in the resolution of the puzzle of
the magnetotransport anomalies in the high-$T_c$ superconductors: the observed {\em
angle-dependent elastic scattering} and the {\em angle-independent T-linear (MFL)
inelastic scattering}. The latter is obtained as scattering off a scale
invariant fluctuation spectrum \cite{mfl}. Indeed, essentially
all of the principal normal state anomalies in the high-$T_c$ superconductors near
optimal composition have now been shown to follow from the MFL fluctuation spectrum.
Furthermore, since the interaction vertices which give the normal state scattering
rates also contribute to Cooper pairing, we expect that these fluctuations, which have the
right magnitude of cut-off frequency
$\omega_c$ and coupling constant
$\lambda$, mediate high-$T_c$ superconductivity.

The authors thank the Aspen Center for Physics where part of this work
was carried out. EA was supported in part
by NSF grant DMR99-76665.

\end{document}